\documentclass[12pt, oneside, a4paper]{article}
\usepackage{enumitem}
\usepackage{amsmath}
\usepackage{amssymb}
\usepackage{amsthm}
\usepackage{bbm}
\usepackage{amsfonts}
\usepackage{mathtools}
\usepackage{tikz-cd}
\usetikzlibrary{shapes,arrows}
\mathtoolsset{showonlyrefs}

\DeclareMathOperator*{\argmax}{arg\,max}

\usepackage{etoolbox}

\usepackage[nottoc,notlot,notlof]{tocbibind}

\usepackage{listings}
\usepackage{color}
\definecolor{mygreen}{RGB}{28,172,0}
\definecolor{mylilas}{RGB}{170,55,241}

\newtheorem*{theorem*}{Theorem}
\newtheorem*{remark*}{Remark}

\usepackage{amsthm}
\theoremstyle{remark}

\usepackage{xpatch}

\makeatletter
\xpatchcmd{\@thm}{\thm@headpunct{.}}{\thm@headpunct{}}{}{}
\makeatother

\usepackage{hyperref}
\usepackage[british]{babel}
\usepackage[justification=centering]{caption}
\usepackage{subcaption}
\usepackage{graphicx}
\usepackage{stackengine}
\usepackage{float}
\usepackage{longtable}
\usepackage{tabularx}
\newcolumntype{Y}{>{\centering\arraybackslash}X}
\usepackage{titlesec}
\titleformat{\chapter}[display]
{\normalfont\huge\bfseries}{}{0pt}{\Huge}
\titlespacing*{\chapter} {0pt}{20pt}{40pt}
\usepackage[thinlines]{easytable}
\usepackage{array}

\usepackage[toc]{appendix}

\usepackage{authblk}

\title{Real-time tomography-based Bayesian inference from TCV bolometry data}
\author[1]{D. Hamm\thanks{Corresponding author: daniele.hamm@epfl.ch}}
\author[1]{C. Theiler}
\author[1]{L. Simons}
\author[1]{B. P. Duval}
\author[1]{U. Sheikh}
\author[1]{\\the TCV team\thanks{See author list of \cite{duval_experimental_2024}.}}
\affil[1]{EPFL-SPC, CH-1015 Lausanne, Switzerland}
\date{}                    
\begin{document}

\maketitle
\vspace{-1cm}
\begin{abstract}
Radiated power information is crucial to diagnose and optimize the performance of fusion plasmas.
Traditionally, at the TCV tokamak, radiated power analysis has only ever been possible following plasma discharge termination. However, recently, TCV bolometer data have become available in real-time. This offers the opportunity of integrating the radiated power information into the TCV plasma control system. In this work, we propose a novel real-time tomography-based Bayesian technique allowing estimation of the power radiated from user-defined regions of interest in the plasma.
The real-time estimates are obtained as computationally cheap linear combinations of bolometer 
measurements, using pre-computed coefficients that are optimized for the specific discharge planned. This method is not, thus, trained on a set of synthetic or tomographically reconstructed emissivity profiles.
We detail the derivation of the technique and show its equivalence to traditional tomographic estimates under suitable conditions. 
We then demonstrate that this technique enables accurate real-time estimation of the total, core, divertor and main chamber radiated power, by its application to a representative and heterogeneous set of TCV discharges.
Finally, we discuss the robustness of the technique to faulty detectors, showing that simple precautions allow safe handling of many common issues. The computational routines implementing the described technique are provided as open-source code.
\end{abstract}

\newpage

\section{Introduction}\label{introduction}
Radiation is one of the main energy dissipation processes in fusion plasmas. Therefore, its investigation is crucial for the understanding, development and control of operational scenarios.
In fusion devices, the radiated power is often measured by diagnostics composed of multiple foils, called bolometers \cite{ingesson_chapter_2008}. These detectors measure the total power radiated by the plasma integrated over their fields-of-view. Multichannel bolometer systems are most often designed to probe the emissivity across a poloidal plasma cross-section.
Spatial resolution is obtained by designing fan-shaped foil configurations that acquire views at different angles.
These measurements can then be used to estimate the two-dimensional poloidal emissivity profile by means of tomographic reconstruction \cite{ingesson_chapter_2008}, providing the poloidal radiation distribution.
In particular, this allows the inference of quantities such as the total power radiated by the plasma and the power radiated by specific regions of interest, e.g., the core (inside the separatrix), divertor (below the X-point) and main chamber (above the X-point).\\
Tomographic reconstruction is widely regarded as the most reliable method to estimate the radiated power, with its estimates used as benchmarks for any other developed method.
Most tomographic techniques, however, are based on iterative schemes \cite{ingesson_chapter_2008, anton_x-ray_1996, craciunescu_maximum_2008, hamm_tomography_2025} that are too computationally expensive for real-time applications \cite{murari_control_2024}. Their computational complexity can also be limiting when processing large amounts of data.
Alternative tomographic approaches, such as Gaussian Process Tomography \cite{li_bayesian_2013} and deep learning \cite{ferreira_deep_2019}, have been proposed in an effort towards real-time capability. Still, important challenges remain: the real-time capability of Gaussian process Tomography is currently restricted to priors that do not use magnetic equilibrium information and is hindered by the necessity of tuning its hyperparameters \cite{moser_gaussian_2022}; deep learning approaches critically depend on the training data \cite{mlynar_current_2019} and their deployment for real-time applications requires specialized hardware \cite{ferreira_monitoring_2021}.\\
To overcome these shortcomings, many computationally efficient alternatives for the estimation of the total/core/divertor/main chamber radiated power have been proposed. 
Some approaches approximate the target quantity by a linear combination of (a subset of) bolometer measurements: the linear combination coefficients may be determined from geometric considerations \cite{ingesson_l_c_comparison_1999}, or learned to optimize model fit based on a set of tomographically inverted experimental data \cite{maraschek_real-time_1998, kallenbach_optimized_2012} or synthetic emissivity profiles \cite{van_de_giessen_development_2021, partesotti_improved_2024}. Other approaches use neural networks that take bolometer data as input and are trained on experimental data \cite{ferreira_monitoring_2021, barana_neural_2002}, while others implement an inversion on a very coarse grid of macropixels \cite{murari_control_2024, rossi_hybrid_2024}, trading spatial for temporal resolution.\\
Real-time capable techniques have many applications in plasma control: estimates of the total \cite{pautasso_-line_2002} and core \cite{ferreira_deep_2019, pau_first_2018} radiated power can be used for disruption prediction and mitigation triggering;
monitoring the total \cite{valcarcel_jet_2014} and core \cite{anand_model-based_2021} radiated power is crucial for first wall heat load mitigation systems; control of the divertor radiated power \cite{kallenbach_optimized_2012, eldon_advances_2019}, main chamber radiated power \cite{kallenbach_optimized_2012}, and of the total and core radiating fractions \cite{eldon_characterization_2024} can be used to optimize scenarios and power exhaust handling. Importantly, real-time control of the radiated power will be critical for all large scale fusion devices such as SPARC \cite{li_development_2024} and ITER \cite{guillemaut_real-time_2017}.\\
The real-time capable techniques described above are designed as computationally efficient surrogates of traditional tomographic estimates. They are either trained on tomographic data or benchmarked against them.
Moreover, all the techniques learning a deterministic mapping from bolometer measurements to the quantity of interest suffer from the same drawback. The estimation accuracy will inevitably depend on the plasma scenario and, in particular, its shape and position in the machine vessel. This is true both for mappings based on geometric considerations \cite{ingesson_l_c_comparison_1999, ingesson_radiation_2003} and for statistical approaches that learn a set of weights. In this second case, the training dataset choice is crucial, forcing to either retrain the technique for changing scenarios \cite{maraschek_real-time_1998} or to accept a trade-off between generality and accuracy \cite{partesotti_improved_2024}. We also remark that none of the techniques discussed so far allow uncertainty quantification.\\
In this work, we address these limitations by proposing a novel Bayesian technique for real-time estimation of the radiated power. The technique is tomography-based, directly approximating tomographic estimates by construction rather than designing/learning an approximate mapping. The technique operates on standard inversion pixel grids, achieving real-time capability without sacrificing spatial resolution. The estimates are obtained by a linear combination of bolometer measurements, using coefficients that are not fixed but evaluated for the planned plasma discharge, thus inherently accounting for different plasma scenarios and not requiring any training procedure. Thanks to its Bayesian nature, the technique also provides real-time uncertainty estimates.\\
In this work, we use our technique to infer radiated power information from bolometry data from the TCV tokamak \cite{duval_experimental_2024}. However, the technique is general and can be used to infer quantities from data acquired by other tomography-capable diagnostics, at TCV and other fusion devices. The only requirement is that the diagnostic data be available in real-time.\\
The main contributions of our paper are the following.
\begin{itemize}
    \item We introduce a novel tomography-based Bayesian technique for real-time estimation of radiated power from bolometry data. The technique allows the estimation of the power radiated from any region of interest, together with the associated uncertainty.
    \item We apply our technique to 50 TCV discharges, representative of the most common magnetic configurations routinely achieved at TCV. We present and analyze the results of this study, showing that total, core, divertor and main chamber radiated power are accurately estimated in real-time.
    \item We demonstrate the robustness of the proposed technique to the presence of faulty channels/measurements, which can be ensured, in the vast majority of cases, by taking into account the diagnostic behavior in the immediately preceding discharges.
    \item We provide open access to the implementation of our technique and the analyses conducted for this paper, to promote reuse by the community and ensure reproducibility of the presented results.
\end{itemize}
The paper is structured as follows. In Section 2, we describe the TCV bolometry diagnostic and the Bayesian approach to the associated tomographic reconstruction problem, with particular focus on the case where the posterior distribution is Gaussian. 
In Section 3, we derive our technique.
 First, we show that, for Gaussian posteriors, the power radiated from any given plasma region can be computed by a linear combination of bolometer measurements, with suitably defined coefficients.
Then, we detail how to leverage the magnetic configuration planned during discharge preparation to pre-compute the said coefficients.
In Section 4, we apply the technique to experimental TCV bolometry data. In particular, we demonstrate that the real-time estimates for total, core, divertor and main chamber radiated power approximate well the post-discharge estimates obtained by state-of-the-art tomographic reconstruction methods.
Finally, in Section 5, we conclude by discussing the obtained results and identify possible areas of improvement for future research.

\section{Tomographic Reconstruction Problem}\label{tomography}
\begin{figure}[t]
\centering
\begin{subfigure}[b]{0.20\textwidth}
\includegraphics[width=\textwidth]{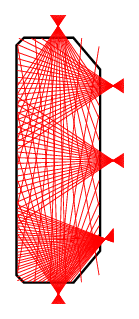}
\end{subfigure}
\hspace{1.cm}
\begin{subfigure}[b]{0.5\textwidth}
\includegraphics[width=\textwidth]{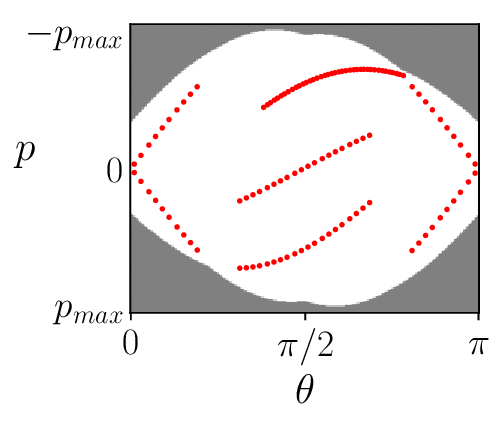}
     \end{subfigure}
\captionsetup{justification=justified,singlelinecheck=true}
\caption{Bolometer diagnostic installed at TCV. On the left, the line of sight configuration (in red) overlaid to a poloidal cross-section of TCV (in black). On the right, the diagnostic coverage in projection space, highlighting the sparse-view nature of the data: each red dot corresponds to a line of sight, parametrized in terms of its steepness $\theta$ and its distance $p$ from the centerpoint of the vessel; the shaded regions represent $(p, \theta)$ combinations corresponding to lines that do not intersect the TCV vessel.}
\label{fig_radcam}
\end{figure}

Nearly all fusion devices are equipped with bolometer cameras. The bolometry system installed at TCV, shown in Figure \ref{fig_radcam}, is part of the RADCAM diagnostic \cite{sheikh_radcam-radiation_2022}.
The RADCAM system consists of carbon coated foil bolometers \cite{sheikh_novel_2016}, SXR diodes and AXUV diodes with overlapping lines of sight for cross-calibration.
The system includes $120$ bolometers, divided in 5 cameras: the top, upper lateral, midlateral and bottom cameras are composed of $20$ detectors, while the lower lateral (divertor) camera is composed of $40$ detectors. All cameras are located in the same poloidal sector of TCV. In the left part of Figure \ref{fig_radcam}, we report the line-of-sight representation of the diagnostic, with each line-of-sight representing the centerline of the volume subtended to the corresponding detector.
The system was designed to optimize poloidal coverage given the technical constraints posed by the port locations. 
The diagnostic is routinely used to tomographically reconstruct the poloidal emissivity profile of the plasma. 
Under the assumption of toroidally symmetric emission, which is customary and typically justified in steady-state operation of tokamaks, the total radiated power can be estimated by toroidally integrating the reconstructed poloidal profile.
Despite the optimized coverage, the recorded data are actually quite sparse: this is clear from the projection space representation of the diagnostic, displayed in the right part of Figure \ref{fig_radcam}. The sparse-view, limited-angle nature of the data makes the resulting tomographic reconstruction problem challenging \cite{hamm_tomography_2025}.\\
The Bayesian approach is particularly well-suited to address this challenge \cite{hamm_tomography_2025}. In Bayesian terms, the tomographic reconstruction problem can be formalized as follows. Let us consider the ill-posed inverse problem of reconstructing the emissivity profile $\mathbf{x}\in\mathbb{R}^N$, with $N$ the number of pixels used to discretize the emissivity in the poloidal plane, from the $M\ll N$ bolometer measurements $\mathbf{y}\in\mathbb{R}^M$. The relation between $\mathbf{y}$ and $\mathbf{x}$ is given by the equation
\begin{equation}\label{ip_data}
    \mathbf{y} = \mathbf{T}\mathbf{x} + \boldsymbol{\varepsilon},
\end{equation}
where $\mathbf{T}\in\mathbb{R}^{M\times N}$ is the so-called geometry matrix \cite{ingesson_chapter_2008} modeling the acquisition system and $\boldsymbol{\varepsilon}\in\mathbb{R}^M$ is the noise affecting the measurements.
In Bayesian inference, a statistical model of the noise is used to define the likelihood distribution $p(\mathbf{y}\vert\mathbf{x})$, while the assumed properties of $\mathbf{x}$ are encoded in the prior distribution $p(\mathbf{x})$. Then, our knowledge about the unknown emissivity profile given the observed tomographic data is expressed by the posterior distribution $p(\mathbf{x}\vert\mathbf{y})$. By Bayes' theorem, $p(\mathbf{x}\vert\mathbf{y})$ satisfies
\begin{equation}\label{bayes_formula}
    p(\mathbf{x}\vert \mathbf{y}) \propto p(\mathbf{y} \vert \mathbf{x}) p(\mathbf{x})\,.
\end{equation}
Bayesian inference consists in extracting information from $p(\mathbf{x}\vert\mathbf{y})$. For example, the most credible solution to the inverse problem is given by the maximum a posteriori (MAP) estimator, which is given by
\begin{equation}\label{MAP}
\mathbf{x}_{_{MAP}}=\argmax_{\mathbf{x}\in\mathbb{R}^N}\; p(\mathbf{x}\,\vert \mathbf{y}). 
\end{equation}
In this work, we always consider noise models such that the noise $\varepsilon_m$ affecting each measurement $y_m$ is independent of the noise affecting the other channels and follows a Gaussian distribution with zero mean and variance $\eta_m^2$, i.e., $\varepsilon_i\sim\mathcal{N}(0,\eta_m^2)$. Under this assumption, the likelihood is Gaussian, with
\begin{equation}\label{gaussian_likelihood}
    \mathbf{y}\vert\mathbf{x} \sim \mathcal{N}(\mathbf{T}\mathbf{x}, \mathbf{I}_{\boldsymbol{\eta}^2}), \;\; p(\mathbf{y}\,\vert\,\mathbf{x}) \propto \exp\Big(-\frac{1}{2}(\mathbf{y}-\mathbf{T}\mathbf{x})^T\,\mathbf{I}_{\boldsymbol{\eta}^{-2}}\,(\mathbf{y}-\mathbf{T}\mathbf{x})\Big),
\end{equation}
where $\mathbf{I}_{\boldsymbol{\eta}^2},\;\mathbf{I}_{\boldsymbol{\eta}^{-2}}$ are diagonal matrices such that $(\mathbf{I}_{\boldsymbol{\eta}^2})_{mm}=\eta_m^2$, $(\mathbf{I}_{\boldsymbol{\eta}^{-2}})_{mm}=\eta_m^{-2}$, for $m=1,\ldots,M$.  
Moreover, we focus on the case where the posterior itself is Gaussian. As discussed in \cite{hamm_tomography_2025}, this situation is not exclusive to the framework commonly referred to as Gaussian process tomography \cite{li_bayesian_2013, moser_gaussian_2022}, but arises from the larger class of Gaussian smoothness priors \cite{kaipio_statistical_2005}. In particular, it arises for the class of diffusion priors described in \cite{hamm_tomography_2025} that we use in this work, for which it holds
\begin{equation}\label{eq_diffusion_prior}
p(\mathbf{x})\propto \exp\!\!\,\Big(\!\!-\frac{1}{2} \big\Vert \sqrt{\mathbf{D}(\psi,\alpha)}\boldsymbol{\nabla}\mathbf{x} \big\Vert_2^2 \Big),
\end{equation}
where $\mathbf{D}(\psi,\alpha)$ is a diffusion tensor that depends on the poloidal flux function $\psi$ and with anisotropic parameter $\alpha$. The dependence on $\psi$ encodes magnetic field information in the prior. If $\alpha=1$, magnetic information is not used and Eq.\eqref{eq_diffusion_prior} promotes isotropic smoothing, while $\alpha<1$ anisotropically favors smoothness along, rather than across, flux surfaces. The effect of this prior is similar \cite{hamm_tomography_2025} to that of anisotropic regularization techniques commonly used in plasma tomography \cite{ingesson_chapter_2008}.\\
Gaussian posteriors are popular because their mean and covariance matrix can be expressed in closed-form.
For the likelihood in Eq.\eqref{gaussian_likelihood} and prior in Eq.\eqref{eq_diffusion_prior}, the resulting Gaussian posterior \cite{hamm_tomography_2025} is given by $\mathbf{x}\vert\mathbf{y}\sim\mathcal{N}(\boldsymbol{\mu}_{post},\,\boldsymbol{\Sigma}_{post})$, with
\begin{equation}\label{eqs_posterior}
    \boldsymbol{\mu}_{post} = \boldsymbol{\Sigma}_{post} \mathbf{T}^T\,\mathbf{I}_{\boldsymbol{\eta}^{-2}}\,\mathbf{y},\quad \boldsymbol{\Sigma}_{post} = \big(\mathbf{T}^T\,\mathbf{I}_{\boldsymbol{\eta}^{-2}}\,\mathbf{T}\,+\, \lambda\, \boldsymbol{\nabla}^T\,\mathbf{D}(\psi,\alpha)\boldsymbol{\nabla} \big)^{-1},
\end{equation}
where $\lambda>0$ is a prior hyperparameter that plays the role of regularization parameter \cite{hamm_tomography_2025}. Since the posterior is Gaussian, the MAP estimate defined in Eq.\eqref{MAP} coincides with the posterior mean $\boldsymbol{\mu}_{post}$ in Eq.\eqref{eqs_posterior}$\,$: the mode and mean of a Gaussian distribution coincide.
The matrix
\begin{equation}\label{eq_matrix_A}
\mathbf{A}=\boldsymbol{\Sigma}_{post}\mathbf{T}^T\,\mathbf{I}_{\boldsymbol{\eta}^{-2}}\in\mathbb{R}^{N\times M}
\end{equation}
defines the linear transformation mapping the tomographic data to the poloidal plane, with $\boldsymbol{\mu}_{post}=\mathbf{A}\mathbf{y}$.
If $\eta_m=\eta$ for $m=1,\ldots,M$, i.e., if the noise level is the same for all channels, Eq.\eqref{eqs_posterior} simplifies to
\begin{equation}\label{eqs_posterior_flatnoise}
    \boldsymbol{\mu}_{post} = \eta^{-2}\boldsymbol{\Sigma}_{post} \mathbf{T}^T\,\mathbf{y},\quad \boldsymbol{\Sigma}_{post} = \big(\eta^{-2}\,\mathbf{T}^T\,\mathbf{T}\,+\, \lambda\, \boldsymbol{\nabla}^T\,\mathbf{D}(\psi,\alpha)\boldsymbol{\nabla} \big)^{-1}\;.
\end{equation}

\section{Radiated power estimation: a real-time\\ tomography-based Bayesian approach}\label{method}
\begin{figure}[t]
\vspace{-0.2cm}
\begin{subfigure}[b]{\textwidth}
\!\!\includegraphics[width=\textwidth]{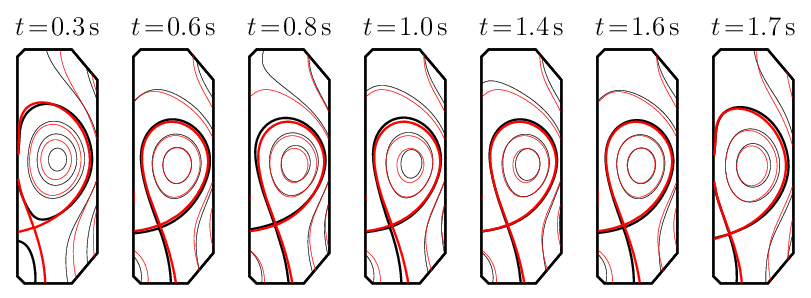}
\end{subfigure}
\captionsetup{justification=raggedright,singlelinecheck=true}
\caption{FBT vs LIUQE magnetic equilibria, for TCV discharge $\#85270$. In red, FBT equilibria, in black, LIUQE equilibria.}
\label{fig_fbt_vs_liuqe}
\vspace{-0.5cm}
\end{figure}

In this section, we introduce our technique for real-time estimation of the power radiated by arbitrary regions of interest of the plasma. First, let us consider the problem of estimating the total radiated power $P_{rad}^{\,tot}$. We model our knowledge about the poloidal emissivity profile $\mathbf{x}$ using the posterior described in Section \ref{tomography}. Given a profile $\mathbf{x}$, the associated total radiated power $P_{rad}^{\,tot}(\mathbf{x})$ is obtained by integrating $\mathbf{x}$ both poloidally and toroidally. In the discrete pixel-based representation used in this work, this amounts to computing a weighted sum of pixel values. We consider a uniform rectangular pixel grid in the poloidal plane, with $\Delta Z$ and $\Delta R$ the vertical and radial size of each pixel, respectively. Denoting by $R_i$ the radial location of the center of the $i$th pixel, $P_{rad}^{\,tot}(\mathbf{x})$ is given by
\begin{equation}\label{eq_prad}
\begin{aligned}
P_{rad}^{\,tot}(\mathbf{x}) &= \sum_{i=1}^{N} \underbrace{2\pi \,R_i\, \Delta Z \Delta R}_\text{$\eqqcolon\,b_i$}\, x_i\;\\
&= \mathbf{b}^T\mathbf{x},
\end{aligned}
\end{equation}
with $\mathbf{b}\in\mathbb{R}^{N\times1}$; the factors $2\pi R_i$ and $\Delta Z \Delta R$ account for toroidal and poloidal integration, respectively.
In particular, Eq.\eqref{eq_prad} shows that the radiated power is a linear combination of pixel values: thus, it is given by a linear transformation of the vector $\mathbf{x}\in\mathbb{R}^N$. The posterior distribution is given by $\mathbf{x}\vert\mathbf{y}\sim\mathcal{N}(\boldsymbol{\mu}_{post},\,\boldsymbol{\Sigma}_{post})$. As a consequence, due to a well-known property of multivariate Gaussians (see e.g. \cite[Ch. 3]{mathai_multivariate_2022}), the posterior distribution of the total radiated power is a univariate Gaussian satisfying
\begin{equation}\label{eq_prad_posterior}
P_{rad,\,post}^{\,tot} \sim \mathcal{N}(\mathbf{b}^T\boldsymbol{\mu}_{post},\;\mathbf{b}^T\,\boldsymbol{\Sigma}_{post}\,\mathbf{b})\;.
\end{equation}
The Gaussianity of the posterior $\mathbf{x}\vert\mathbf{y}$ thus implies the Gaussianity of the radiated power posterior estimate. The most credible (MAP) estimate of the total radiated power, which we denote by $\widehat{P}_{rad}^{\,tot}$, is the mean of $P_{rad,\,post}^{\,tot}$ in Eq.\eqref{eq_prad_posterior}. Crucially, since $\boldsymbol{\mu}_{post}=\mathbf{A}\mathbf{y}$ with $\mathbf{A}$ defined in Eq.\eqref{eq_matrix_A}, $\widehat{P}_{rad}^{\,tot}$ is given by 
\begin{equation}\label{eq_coefficients}
\begin{aligned}
\widehat{P}_{rad}^{\,tot} &= \sum_{i=1}^{N} b_i (\boldsymbol{\mu}_{post})_i\\
&= \sum_{i=1}^{N} b_i \Big(\sum_{j=1}^MA_{ij}\,y_j\Big)\\
&= \sum_{j=1}^M\,\underbrace{\sum_{i=1}^{N} b_i\,A_{ij}}_\text{$\eqqcolon\;\beta_j$}\;y_j\;\\[-1.5ex]
&= \sum_{j=1}^M\,\beta_j\;y_j\;.\\
\end{aligned}
\end{equation}
Therefore, by inverting the order of the summations over the pixels and over the detectors as shown in Eq.\eqref{eq_coefficients}, $\widehat{P}_{rad}^{\,tot}$ can be written as a linear combination of tomographic measurements, with coefficients $\boldsymbol{\beta}\in\mathbb{R}^M$ ultimately given by
\begin{equation}\label{eq_beta}
\begin{aligned}
    \beta_j = \sum_{i=1}^N2\pi R_i\Delta Z \Delta R \Big(\big(\mathbf{T}^T\,\mathbf{I}_{\boldsymbol{\eta}^{-2}}\,\mathbf{T}\,+\, \lambda\, \boldsymbol{\nabla}^T\,\mathbf{D}(\psi,\alpha)&\boldsymbol{\nabla} \big)^{-1}\,\mathbf{T}^T\,\mathbf{I}_{\boldsymbol{\eta}^{-2}}\Big)_{ij}\;,\\&j=1,\ldots,M.
    \end{aligned}
\end{equation}
Moreover, since the posterior covariance matrix $\boldsymbol{\Sigma}_{post}$ is positive definite \cite[Th. 3.10]{kaipio_statistical_2005}, the variance of  $P_{rad,\,post}^{tot}$ in Eq.\eqref{eq_prad_posterior}, which we will denote by $\sigma_{tot}^2$, is larger than 0. This derivation shows that, for Gaussian posterior models, estimating $P_{rad}^{\,tot}$ by the simple and computationally efficient expression in Eq.\eqref{eq_coefficients} is perfectly equivalent to
integrating poloidally and toroidally the usual two-dimensional MAP tomographic reconstruction. \emph{Thus, by pre-computing the coefficients $\boldsymbol{\beta}$ in Eq.\eqref{eq_beta} before the plasma discharge, $P_{rad}^{\,tot}$ can be estimated in real-time by a linear combination of bolometer measurements $\mathbf{y}$; pre-computing the variance $\sigma_{tot}^2=\mathbf{b}^T\,\boldsymbol{\Sigma}_{post}\mathbf{b}$ from Eq.\eqref{eq_prad_posterior} additionally provides a real-time estimate of the uncertainty associated with $\widehat{P}_{rad}^{\,tot}$.} Importantly, the same approach can be applied to any region of interest $\mathcal{R}$ of the plasma (core, divertor, main chamber, $\ldots$). Indeed, let $I=\{i\in\{1,\ldots,N\}\;\mathrm{such\;that}\;i\in\mathcal{R}\}$ be the set containing the indices of all the pixels belonging to $\mathcal{R}\;$: then, in Eq.\eqref{eq_coefficients}, it suffices to define the $j$th coefficient as $\beta_j=\sum_{i\in I}b_iA_{ij}$ rather than summing over all pixels. Once again, the resulting estimates ($\widehat{P}_{rad}^{\,core}$, $\widehat{P}_{rad}^{\,div}$, $\widehat{P}_{rad}^{\,main}$, $\ldots$) are mathematically equivalent to the estimates computed from the two-dimensional MAP.
\\
This provides a clear path towards real-time tomography-based Bayesian estimation of the power radiated from any plasma region of interest. However, a few critical challenges in the application of this technique remain to be addressed.
The coefficients in Eq.\eqref{eq_beta} depend on the magnetic equilibrium $\psi$, the possibly signal-dependent noise level $\boldsymbol{\eta}$, and the prior hyperparameters $\alpha, \lambda$. We now discuss a strategy for defining/determining these quantities prior to a given discharge.\newline
Let us first consider the $\psi$-dependence of the coefficients $\boldsymbol{\beta}$. This arises in estimators for specific plasma regions $\mathcal{R}$ and/or if $\alpha<1$ in the prior in Eq.\eqref{eq_diffusion_prior}. As already remarked, if $\alpha<1$, this prior promotes smoothness along flux surfaces. In offline tomography analysis, such priors are defined using the plasma equilibrium reconstructed by dedicated codes. At TCV, for example, $\psi$ is reconstructed from magnetic measurements by using the LIUQE \cite{hofmann_tokamak_1988, Moret-FusEngDes2015-LIUQE-MATLAB} code. Although a real-time implementation of LIUQE exists \cite{Moret-FusEngDes2015-LIUQE-MATLAB}, the computation of $\boldsymbol{\beta}$ in Eq.\eqref{eq_beta} features expensive matrix operations and is not efficient enough to leverage its output: the coefficients must thus be pre-computed. Therefore, instead of employing the LIUQE reconstruction $\psi_{_{LIUQE}}$, we define the coefficients based on the plasma equilibrium $\psi_{_{FBT}}$ computed by FBT \cite{Hofmann-CPC-1988-FBT}, a free-boundary tokamak equilibrium code. FBT is routinely used at TCV during discharge preparation: given a desired plasma shape and a set of constraints, it computes a plasma equilibrium $\psi_{_{FBT}}$ and the poloidal field coil currents required to achieve it. In practice, FBT outputs a sequence of times within the discharge $\{t_{_{FBT}}^k\}_{k=1}^K$ and the corresponding plasma shapes expected at those times $\{\psi_{_{FBT}}^k=\psi_{_{FBT}}(t_{_{FBT}}^k)\}_{k=1}^K$.
Typically, $K$ is between $20$ and $30$. To leverage this information, we compute $K$ different sets of coefficients $\{\boldsymbol{\beta}_{_{FBT}}^k\}_{k=1}^K$. Then, for real-time estimation at time $t$, we pick the coefficients corresponding to the closest FBT time $t_{_{FBT}}^k$, using the $k$th set of coefficients if $t\in[(t_{_{FBT}}^{k-1}+t_{_{FBT}}^k)/2,\;(t_{_{FBT}}^{k}+t_{_{FBT}}^{k+1})/2\;]$. Computing $\{\boldsymbol{\beta}_{_{FBT}}^k\}_{k=1}^K$, while not feasible in real-time, can be done during discharge preparation. Moreover, if the main interest is in the flat-top operating phase, as is often the case, the FBT equilibria $\psi_{_{FBT}}^k$ associated to the breakdown, ramp-up and ramp-down phases can be neglected, reducing the number of coefficients to be computed and deployed. The LIUQE reconstructions, that are much closer to the actual plasma equilibria, typically differ from the FBT calculations. Nevertheless, in most cases, they are quite similar. In Figure \ref{fig_fbt_vs_liuqe}, we show a comparison of the equilibria from the two codes for a few selected FBT times of TCV discharge $\#85270$.
This case is fairly typical and allows an appreciation of some of the common differences: the limited-diverted plasma transition  may occur at slightly different times (see $t=0.3\,\mathrm{s}$), the last closed magnetic flux surface may be partially deformed or displaced (see $t=0.8\,\mathrm{s}, 1.0\,\mathrm{s}$) or the (outer) divertor leg may deviate from its intended location (see all times $t\geq0.3\,\mathrm{s}$). Despite these differences, as we will show in Section \ref{results}, calculating the coefficients using the FBT equilibria is sufficient for accurate real-time estimation. It is worth noticing that recently proposed improved shape control tools \cite{mele_design_2025} are expected to reduce even further any disagreements between the desired and experimentally observed equilibria.\\
Let us now consider the dependence of $\boldsymbol{\beta}$ on the noise level $\boldsymbol{\eta}$ and the prior hyperparameters $\alpha, \lambda$. The two-dimensional tomographic reconstructions, in the sparse-view regime characterizing plasma tomography, can be quite sensitive to these parameters \cite{hamm_tomography_2025}. However, careful hyperparameter tuning is far less crucial for estimating the power radiated from large plasma regions \cite{hamm_tomography_2025, ingesson_l_c_comparison_1999, david_optimization_2021}. In Appendix \ref{appendix_hyperparam}, we discuss a hyperparameter study conducted on $10^3$ physically realistic synthetic emissivity profiles. Its results indicate that the radiated power, upon proper normalization of the input tomographic data, can be accurately estimated by a posterior model assuming a signal-independent fixed noise level $\eta$ (even when the data are actually affected by signal-dependent noise) and fixed ``average'' values for the prior hyperparameters $\alpha, \lambda$.
This key observation addresses the last remaining challenge to the pre-computation of the coefficients in Eq.\eqref{eq_beta}.

\section{Results: application to TCV bolometry data}\label{results}
\begin{figure}[H]   
\vspace{-0.3cm}
\captionsetup[subfigure]{oneside, skip=-4pt, margin={-1.3cm,0cm}}
\begin{subfigure}[]{\textwidth}
\centering
\includegraphics[width=13.5cm,height=5cm]{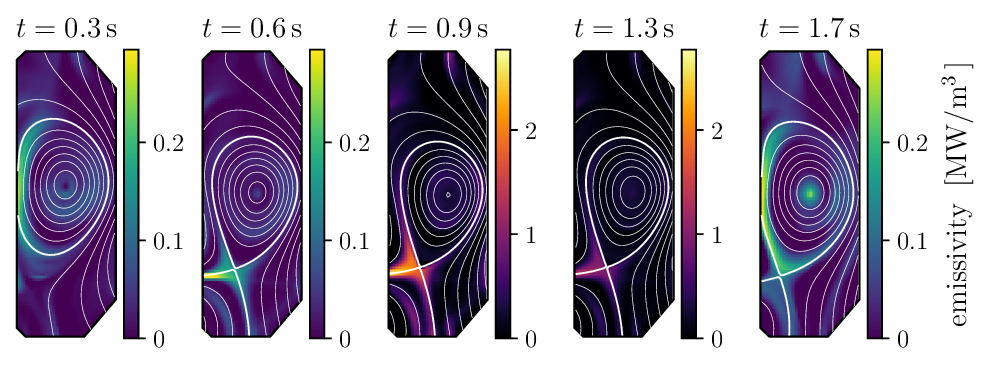}
\caption{}
\label{fig_shot_inversions}
\end{subfigure}
\\
\vspace{0.15cm}
\\
\captionsetup[subfigure]{oneside, skip=0pt, margin={-0.cm,0cm}}
\begin{subfigure}[]{0.15\textwidth}
\centering
\raisebox{15pt}{
\includegraphics[width=1.84cm,height=4.8cm]{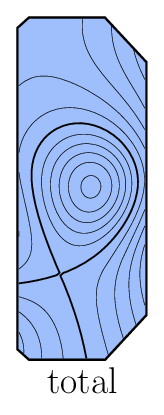}
}
\caption{}
\label{fig_shot_tot_mask}
\end{subfigure}
\hspace{0.cm}
\begin{subfigure}[]{0.15\textwidth}
\centering
\raisebox{14pt}{
\includegraphics[width=1.84cm,height=4.8cm]{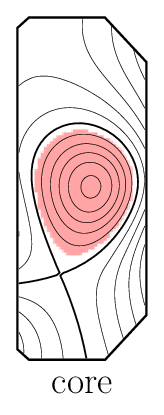}
}
\caption{}
\label{fig_shot_core_mask}
\end{subfigure}
\hspace{0.2cm}
\captionsetup[subfigure]{oneside, skip=0pt,margin={0.7cm,0cm}}
\begin{subfigure}[]{0.7\textwidth}
\includegraphics[width=9cm,height=5.5cm]{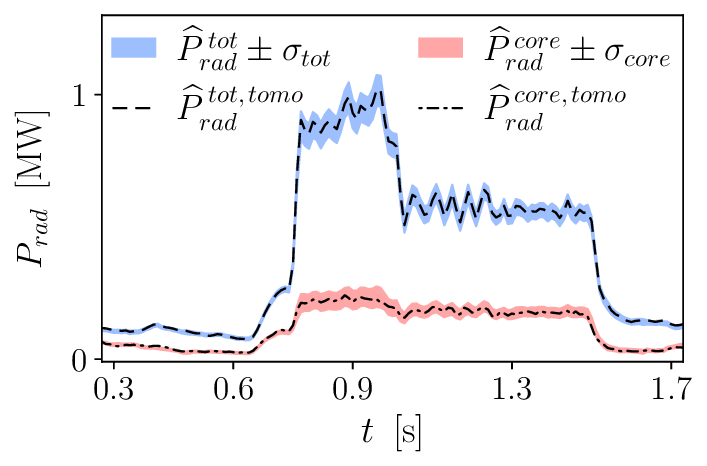}
\caption{}
\label{fig_shot_prad_tot_core}
\end{subfigure}
\\
\vspace{0.15cm}
\\
\captionsetup[subfigure]{oneside, skip=0pt, margin={-0.cm,0cm}}
\begin{subfigure}[]{0.15\textwidth}
\centering
\raisebox{19pt}{
\includegraphics[width=1.84cm,height=4.8cm]{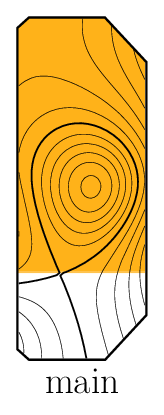}
}
\caption{}
\label{fig_shot_main_mask}
\end{subfigure}
\hspace{0.cm}
\begin{subfigure}[]{0.15\textwidth}
\centering
\raisebox{19pt}{
\includegraphics[width=1.84cm,height=4.8cm]{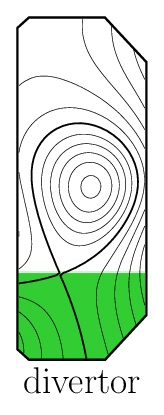}
}
\caption{}
\label{fig_shot_div_mask}
\end{subfigure}
\hspace{0.2cm}
\captionsetup[subfigure]{oneside, skip=0pt,margin={0.7cm,0cm}}
\begin{subfigure}[]{0.7\textwidth}
\includegraphics[width=9cm,height=5.5cm]
{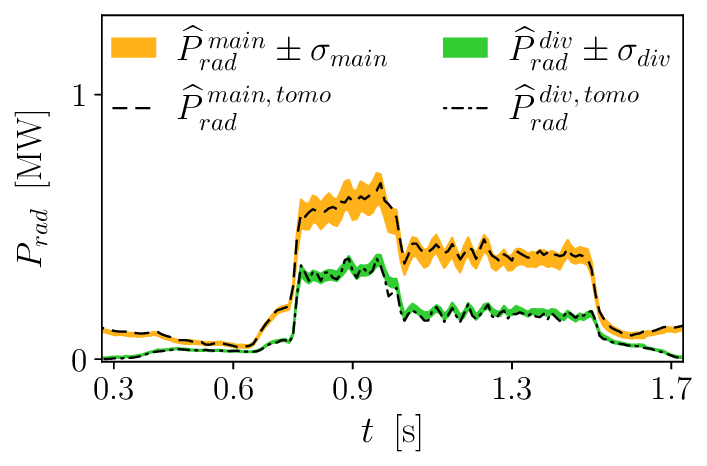}
\caption{}
\label{fig_shot_prad_main_div}
\end{subfigure}
\\
\vspace{-0.7cm}
\\
\captionsetup{justification=justified,singlelinecheck=true}
\caption{Radiated power estimation: results of application to discharge $\#85270$. In Fig. \ref{fig_shot_inversions}, the post-discharge tomographic inversions at 5 reference times; due to the time-varying emissivity levels, two different colormaps and colorbars are used for visualization purposes.
In Figs.\ref{fig_shot_tot_mask}, \ref{fig_shot_core_mask}, \ref{fig_shot_main_mask} and \ref{fig_shot_div_mask}, the masks used to estimate the total, core, main chamber and divertor radiated power (for $t_{_{FBT}}=1\,\mathrm{s}$). In Figs.\ref{fig_shot_prad_tot_core} and \ref{fig_shot_prad_main_div}, the radiated powers $\widehat{P}_{rad}^{\,tomo}$ estimated by tomographic reconstruction and the real-time estimates $\widehat{P}_{rad}\pm\sigma$.}
\label{fig_shot}
\end{figure}

\begin{figure}[]
\captionsetup[subfigure]{oneside, skip=0pt,margin={-0.1cm,0cm}}
\begin{subfigure}[]{0.23\textwidth}
\centering
\includegraphics[width=3.08cm,height=4.8cm]{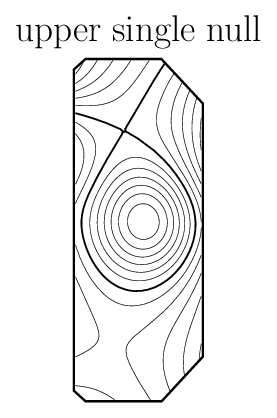}
\caption{}
\label{usn_config}
\end{subfigure}
\hspace{-0.1cm}
\captionsetup[subfigure]{oneside, skip=0pt,margin={0.1cm,0cm}}
\begin{subfigure}[]{0.23\textwidth}
\centering
\includegraphics[width=3.9cm,height=4.8cm]{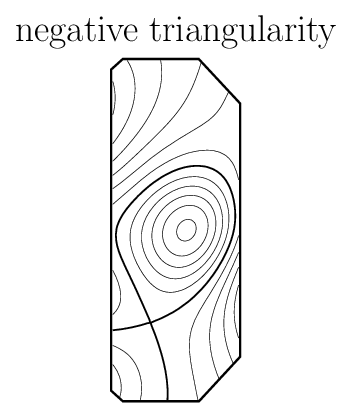}
\caption{}
\label{nt_config}
\end{subfigure}
\hspace{0.6cm}
\captionsetup[subfigure]{oneside, skip=0pt,margin={0.1cm,0cm}}
\begin{subfigure}[]{0.23\textwidth}
\centering
\includegraphics[width=2.7cm,height=4.8cm]{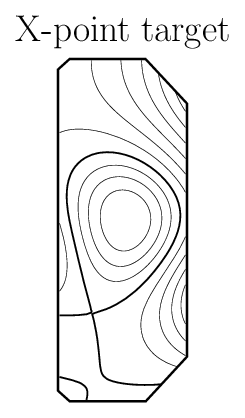}
\caption{}
\label{xpt_config}
\end{subfigure}
\hspace{0.025cm}
\captionsetup[subfigure]{oneside, skip=0pt,margin={0.1cm,0cm}}
\begin{subfigure}[]{0.23\textwidth}
\centering
\includegraphics[width=2.275cm,height=4.8cm]{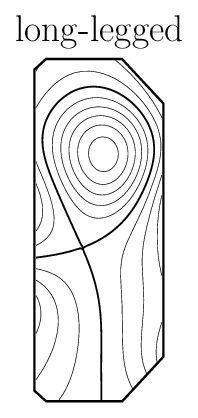}
\caption{}
\label{ll_config}
\end{subfigure}
\\
\vspace{0.3cm}
\\
\captionsetup[subfigure]{oneside, skip=0pt,margin={0.7cm,0cm}}
\begin{subfigure}[]{0.5\textwidth}
\centering
$\!\!\!\!\!$\includegraphics[width=7cm,height=5.5cm]{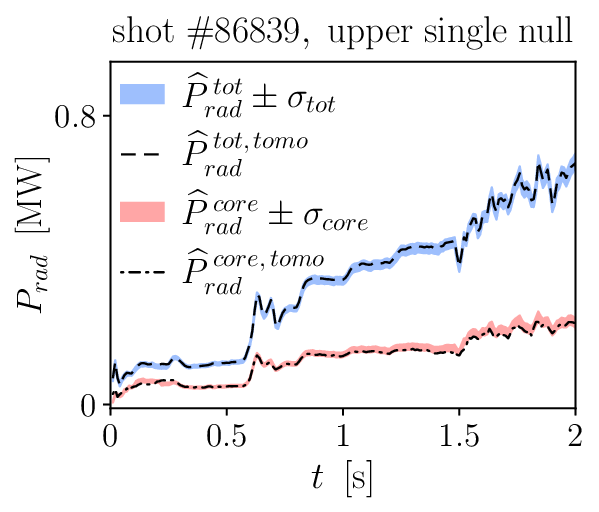}
\caption{}
\label{usn_prad}
\end{subfigure}
\hspace{0.cm}
\captionsetup[subfigure]{oneside, skip=0pt,margin={0.85cm,0cm}}
\begin{subfigure}[]{0.5\textwidth}
\centering
\includegraphics[width=7cm,height=5.5cm]{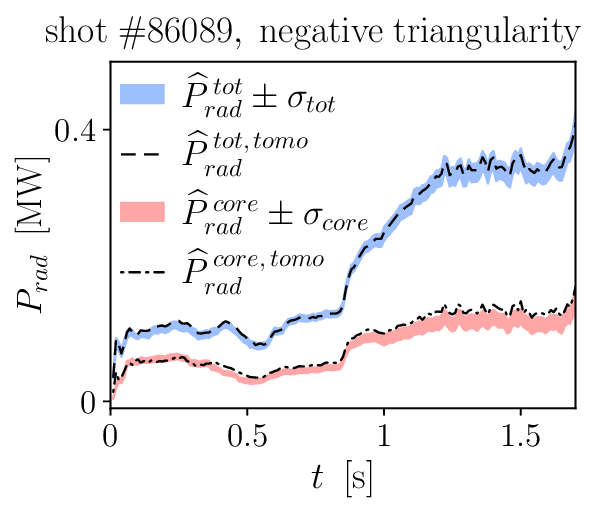}
\caption{}
\label{nt_prad}
\end{subfigure}
\\
\vspace{0.3cm}
\\
\captionsetup[subfigure]{oneside, skip=0pt,margin={0.7cm,0cm}}
\begin{subfigure}[]{0.5\textwidth}
\centering
$\!\!\!\!\!$\includegraphics[width=7cm,height=5.5cm]{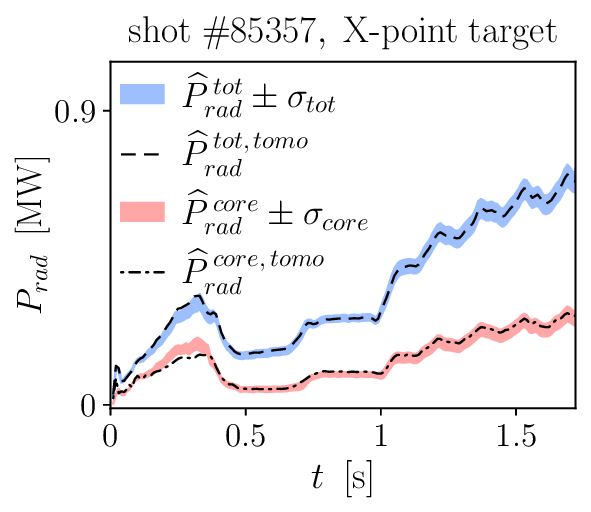}
\caption{}
\label{xpt_prad}
\end{subfigure}
\hspace{0.cm}
\captionsetup[subfigure]{oneside, skip=0pt,margin={0.7cm,0cm}}
\begin{subfigure}[]{0.5\textwidth}
\centering
\includegraphics[width=7cm,height=5.5cm]{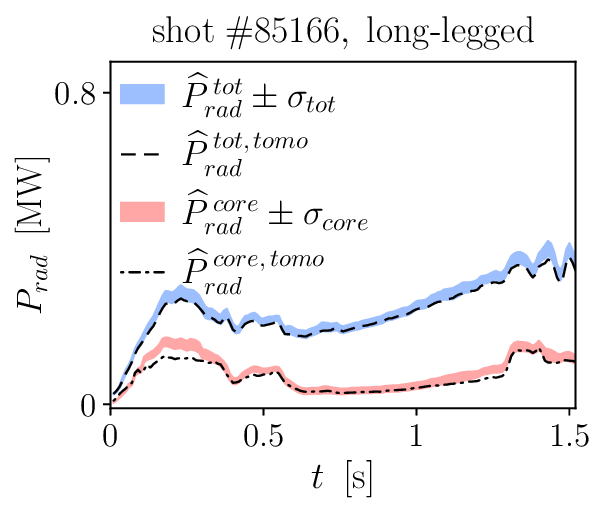}
\caption{}
\label{ll_prad}
\end{subfigure}
\\
\vspace{-0.7cm}
\\
\captionsetup{justification=justified,singlelinecheck=true}
\caption{Radiated power estimation: results of application to discharges $\#86839$, $\#86089$, $\#85357$ and $\#85166$, characterized by different magnetic configurations. In Figs. \ref{usn_config}-\ref{ll_config}, FBT magnetic equilibria at $t\approx1\,\mathrm{s}$ for the four discharges. In Figs. \ref{usn_prad}-\ref{ll_prad}, results for the estimation of total and core radiated power (same plotting conventions of Figs. \ref{fig_shot_tot_mask}-\ref{fig_shot_prad_tot_core}). Results for divertor and main chamber radiated power are not shown to avoid redundancy.}
\label{fig_shots_configs}
\end{figure}

In this section, we apply the technique introduced in Section \ref{method} to experimental TCV data. All the computational routines and data used to generate the presented results are available, in open access form, at the paper GitHub repository \url{https://github.com/dhamm97/real-time-tomo-prad.git}. For the implementation of the Bayesian approach discussed in section \ref{tomography}, we employ the open-source computational imaging framework \texttt{Pyxu}\cite{pyxu-framework}; in particular, the diffusion prior in Eq.\eqref{eq_diffusion_prior} is implemented using the \texttt{Pyxu} plug-in \texttt{pyxu-diffops}\cite{pyxu-diffops}.\\
We study the performance of our technique on 50 TCV discharges, all from the TCV experimental campaign in the first half of 2025. They were selected to be representative of some of the most relevant diverted configurations investigated at TCV. We include 10 examples for each of the following magnetic configurations: lower single null, upper single null, negative triangularity, X-point target and long-legged. Considering different configurations is important, as any technique that is to be routinely deployed in operations must work well regardless of plasma shape. This is especially true at TCV, due to its high plasma shaping flexibility.
More details on the selection process and the numbers of the used discharges are reported in Appendix \ref{shot_numbers}.\\
In this study, we also investigate the robustness of our technique to the failure of one or multiple detectors. Identifying robust strategies to deal with this event is crucial, since detector failure is a rather common occurrence in plasma operation. Failures may lead to specific channels not measuring any signal, measuring signals that are larger/smaller than they should be, or exhibiting unphysical oscillations. At TCV, a procedure, run after every discharge, identifies any channels that were ill-behaved. The corresponding detectors are then excluded from those used for offline tomographic reconstruction. For the 50 examples analyzed in this section, the procedure identified between 3 and 15 bad channels per discharge, with an average of $\approx6$. Clearly, real-time techniques like ours cannot rely on such a post-discharge procedure. We study the performance of our technique for four different channel selection strategies, which we refer to as $S_1$, $S_2$, $S_3$ and $S_4$. For all four, we exclude from the analysis three chords that almost always exhibit erroneous measurements. In $S_1$, all the other $117$ channels are used to compute the coefficients in Eq.\eqref{eq_beta}. In $S_2$, we use only the channels that were determined to be well-behaved by the post-discharge procedure: as such, it cannot be implemented in real-time applications.
The results obtained with $S_2$ are to be compared with those from offline tomographic analysis, to compare the two estimates when using the same channels.
In $S_3$, we use only the channels that were well-behaved for the immediately preceding discharge. This strategy is reasonable, as failure patterns are often consistent over time.
Contrarily to $S_2$, $S_3$ can be implemented in operational settings. By comparing $S_3$ with $S_2$, we can study the influence of unexpected ill-behaved channels on the estimation. In the 50 analyzed discharges, there are four such instances. Finally, in $S_4$, we consider all the channels from $S_3$ except four ones that are often ill-behaved. Comparison of $S_4$ to $S_3$ can be used to probe the sensitivity of our technique to the removal of a few chords. Inspecting the failure patterns of the over $1000$ discharges of the aforementioned 2025 campaign, we find that implementing $S_4$ would have avoided using even a single ill-behaved channel in $\approx97\%$ of the experiments. Therefore, this strategy is expected to effectively prevent the use of ill-behaved channels in most cases.\\
For each of the selected discharges, we estimate the total, core, main chamber and divertor radiated power. To do this, we first compute the post-discharge tomographic estimates $\widehat{P}_{rad}^{\,tomo}$. These estimates use the LIUQE-reconstructed magnetic equilibria and are obtained by the model described in Appendix \ref{appendix_hyperparam}, using only the channels determined to be well-behaved by the post-discharge routine. As discussed in Appendix \ref{appendix_hyperparam}, $\widehat{P}_{rad}^{\,tomo}$ is expected to be rather accurate: we thus use such estimates as benchmark.
Then, we compute the estimates $\widehat{P}_{rad}$ from the FBT-based technique described in Section \ref{method}. Such estimates correspond to what our technique would have provided if deployed in real-time.
We compute them for all four considered channel selection strategies, using the model described in Appendix \ref{appendix_hyperparam}. As discussed in Section \ref{method}, the standard deviation $\sigma$ can be used to quantify the uncertainty associated with $\widehat{P}_{rad}$.
Credible bounds can be obtained, e.g., as $\widehat{P}_{rad}\pm h\,\sigma,$ with $h\in\mathbb{Z}^+$.
We assess the performance of our technique by comparing its predictions $\widehat{P}_{rad}\pm\sigma$ with the post-discharge estimates $\widehat{P}_{rad}^{\,tomo}$.\\
In Figure \ref{fig_shot}, we provide a complete visual representation of our study, for the lower single null discharge $\#85270$. In Figure \ref{fig_shot_inversions}, we display the tomographic reconstructions used to compute the estimates $\widehat{P}_{rad}^{\,tomo}$. The masks defining the regions of interest are shown in Figures \ref{fig_shot_tot_mask}, \ref{fig_shot_core_mask}, \ref{fig_shot_main_mask} and \ref{fig_shot_div_mask} ; the core was
$\!\!\!\!\!\!\!\!$defined as the region corresponding to a normalized effective radius $\leq0.95$ (with other choices possible). In Figures \ref{fig_shot_prad_tot_core} and \ref{fig_shot_prad_main_div}, we show the performance of the real-time technique for the estimation of $P_{rad}^{\,tot}$, $P_{rad}^{\,core}$, $P_{rad}^{\,main}$, $P_{rad}^{\,div}$. These results are obtained for the channel selection strategy $S_3$. From Figures \ref{fig_shot_prad_tot_core} and \ref{fig_shot_prad_main_div}, we see that the real-time estimates predict very accurately the tomographic estimates $\widehat{P}_{rad}^{\,tomo}$. The estimates $\widehat{P}_{rad}$ closely track the evolution and fluctuations of $\widehat{P}_{rad}^{\,tomo}$, with $\widehat{P}_{rad}\pm\sigma$ providing meaningful credible intervals.\\
In Figure \ref{fig_shots_configs}, we show the results obtained for four other analyzed discharges: an upper single null, a negative triangularity, an X-point target and a long-legged configuration. For simplicity, we only show the results for $P_{rad}^{\,tot}$ and $P_{rad}^{\,core}$; the performance for $P_{rad}^{\,main}$ and $P_{rad}^{\,div}$ is comparable. The plots confirm the observations from the lower single null discharge in Figure \ref{fig_shot}. This shows that accurate real-time estimates can be obtained for a wide variety of magnetic configurations.
Even when the estimates exhibit some bias, e.g., for $P_{rad}^{\,core}$ in discharges $\#86089$ (for $t>0.9\,\mathrm{s}$) and $\#85166$ (for $t<0.5\,\mathrm{s}$), the overall trend is correctly captured by $\widehat{P}_{rad}^{\,core}$: therefore, these real-time estimates can provide useful information in control applications.\\ 
To quantify performance, we evaluate the following four metrics for each experiment and each region of interest.
Let $\{\widehat{P}_{rad}^{\,tomo}(t_\tau)\}_\tau^\mathrm{T}$ and $\{\widehat{P}_{rad}(t_\tau)\}_\tau^\mathrm{T}$ be the time traces of the radiated power estimates for a given discharge. 
We consider the RMSE $E=\sqrt{1/\mathrm{T}\sum_{\tau=1}^\mathrm{T}(\,\widehat{P}_{rad}(t_\tau)-\widehat{P}_{rad}^{\,tomo}(t_\tau))^2}$, the signed average relative error $\Delta=1/\mathrm{T}\sum_{\tau=1}^\mathrm{T}(\widehat{P}_{rad}(t_\tau)-\widehat{P}_{rad}^{\,tomo}(t_\tau))/\widehat{P}_{rad}^{\,tomo}(t_\tau)$, the Pearson correlation coefficient $r$ between the time traces, and the average number $n_\sigma$ of standard deviations separating $\widehat{P}_{rad}^{\,tomo}$ from $\widehat{P}_{rad}$, defined as $n_\sigma=1/\mathrm{T}\sum_{\tau=1}^\mathrm{T}\vert\widehat{P}_{rad}(t_\tau)-\widehat{P}_{rad}^{\,tomo}(t_\tau)\vert\,/\,\sigma(t_\tau)$. For the divertor and main chamber, only the time indices corresponding to the diverted phase are considered. These metrics complement each other.
The RMSE and relative error are used to quantify accuracy; they should be analyzed jointly for completeness.
The correlation coefficient $r$ measures the shape similarity of the two time traces. It takes values between $-1$ and $1$, with $r=1$ indicating a perfect linear relationship ($\widehat{P}_{rad}=a_1\,\widehat{P}_{rad}^{\,tomo}+a_2$, $a_1,a_2\in\mathbb{R}$). Values close to $1$ thus imply that the overall trend is well predicted, even in the presence of some bias.
For example, for the previously discussed discharges $\#86089$ and $\#85166$, the correlation coefficients are high despite the bias, with $r_{core}=0.994$ and $0.981$, respectively.
Finally, the average number of standard deviations $n_\sigma$ indicates the factor $h$ needed to obtain operationally useful credible intervals for $\widehat{P}_{rad}^{\,tomo}$ as $\widehat{P}_{rad}\pm h\,\sigma$.\\
\renewcommand{\arraystretch}{1.5}
\begin{table}[]
\scalebox{1.}{
  \begin{tabularx}{0.961\textwidth}{c|c|c|c|c|}
    \cline{2-5}
    &${E}_{{tot}}$ & ${\Delta}_{{tot}}$ & ${r}_{\,tot}$ & ${n}_{\sigma_{tot}}$\\ \hline
    $S_1$&$254.5\;\mathrm{kW}\pm453.8\;\mathrm{kW}$  & $37.9\%\pm56.2\%$ & $0.776\pm0.280$ & $3.2\pm4.6$\\ \hline
    $S_2$&$10.5\;\mathrm{kW}\pm7.8\;\mathrm{kW}$  & $-0.1\%\pm3.0\%$ & $0.992\pm0.022$ & $0.7\pm0.4$\\ \hline
    $S_3$&$10.6\;\mathrm{kW}\pm7.9\;\mathrm{kW}$  & $-0.2\%\pm3.1\%$ & $0.992\pm0.022$ & $0.7\pm0.4$\\ \hline
    $S_4$&$10.5\;\mathrm{kW}\pm7.4\;\mathrm{kW}$  & $0.0\%\pm3.0\%$ & $0.992\pm0.023$ & $0.7\pm0.4$\\ \hline
\end{tabularx}}\\
\vspace{0.4cm}\\
\scalebox{1.}{
  \begin{tabularx}{0.961\textwidth}{c|c|c|c|c|}
    \cline{2-5}
    &${E}_{{core}}$ & ${\Delta}_{{core}}$ & ${r}_{\,core}$ & ${n}_{\sigma_{core}}$\\ \hline
    $S_1$&$126.1\;\mathrm{kW}\pm299.4\;\mathrm{kW}$  & $\phantom{2}5.5\%\pm80.1\%$ & $0.741\pm0.345$ & $2.7\pm4.9$\\ \hline
    $S_2$&$12.6\;\mathrm{kW}\pm5.1\;\mathrm{kW}$  & $6.0\%\pm12.6\%$ & $0.974\pm0.025$ & $1.2\pm0.4$\\ \hline
    $S_3$&$12.6\;\mathrm{kW}\pm5.1\;\mathrm{kW}$  & $5.9\%\pm12.3\%$ & $0.974\pm0.025$ & $1.2\pm0.4$\\ \hline
    $S_4$&$12.5\;\mathrm{kW}\pm5.1\;\mathrm{kW}$  & $6.0\%\pm12.3\%$ & $0.974\pm0.025$ & $1.2\pm0.4$\\ \hline
\end{tabularx}}\\
\vspace{0.4cm}\\
\scalebox{1.}{
  \begin{tabularx}{0.961\textwidth}{c|c|c|c|c|}
    \cline{2-5}
    &${E}_{{div}}$ & ${\Delta}_{{div}}$ & ${r}_{\,div}$ & ${n}_{\sigma_{div}}$\\ \hline
    $S_1$&$\phantom{22}67.9\;\mathrm{kW}\pm89.3\;\mathrm{kW}$  & $10.1\%\pm87.7\%$ & $0.655\pm0.329$ & $2.1\pm1.3$\\ \hline
    $S_2$&$10.4\;\mathrm{kW}\pm12.5\;\mathrm{kW}$  & $7.1\%\pm8.6\%$ & $0.975\pm0.051$ & $1.5\pm1.2$\\ \hline
    $S_3$&$10.6\;\mathrm{kW}\pm12.6\;\mathrm{kW}$  & $6.8\%\pm9.3\%$ & $0.970\pm0.064$ & $1.5\pm1.2$\\ \hline
    $S_4$&$10.6\;\mathrm{kW}\pm12.6\;\mathrm{kW}$  & $7.0\%\pm9.2\%$ & $0.970\pm0.062$ & $1.5\pm1.2$\\ \hline
\end{tabularx}}\\
\vspace{0.4cm}\\
\scalebox{1.}{
  \begin{tabularx}{0.961\textwidth}{c|c|c|c|c|}
    \cline{2-5}
    &${E}_{{main}}$ & ${\Delta}_{{main}}$ & ${r}_{\,main}$ & ${n}_{\sigma_{main}}$\\ \hline
    $S_1$&$257.4\;\mathrm{kW}\pm438.5\;\mathrm{kW}$  & $46.1\%\pm67.7\%$ & $0.762\pm0.291$ & $3.1\pm4.5$\\ \hline
    $S_2$&$10.8\;\mathrm{kW}\pm8.2\;\mathrm{kW}$  & $-1.3\%\pm4.5\%$ & $0.994\pm0.011$ & $0.7\pm0.5$\\ \hline
    $S_3$&$10.9\;\mathrm{kW}\pm8.2\;\mathrm{kW}$  & $-1.3\%\pm4.6\%$ & $0.994\pm0.011$ & $0.8\pm0.5$\\ \hline
    $S_4$&$10.8\;\mathrm{kW}\pm7.9\;\mathrm{kW}$  & $-1.0\%\pm4.6\%$ & $0.994\pm0.011$ & $0.8\pm0.5$\\ \hline
\end{tabularx}}
\captionsetup{justification=raggedright,singlelinecheck=false}
\caption{Error analysis for the estimation of total, core, divertor and main chamber radiated power: results of the study performed on 50 TCV dischar-\\ges. For each discharge, we compute the RMSE $E$, the average relative er-\\ror $\Delta$, the Pearson correlation coefficient $r$ and the average number of stan-\\dard deviations $\sigma$ separating the offline estimate $\widehat{P}_{rad}^{\,tomo}$ from the real-time prediction $\widehat{P}_{rad}$.
We report the average and standard deviation of the met-\\rics for all four investigated channel selection strategies $S_i$, $i=1,2,3,4$.}
\label{table_results}
\end{table}
$\!\!\!\!\!$In Table \ref{table_results}, we report the results of the conducted study. For each channel selection strategy, we include the average and standard deviation of each metric over the 50 TCV discharges analyzed. For $S_1$, the estimates are not accurate nor precise. In particular, the standard deviation of the metrics is high, with performance strongly depending on the number of ill-behaved channels in the discharge. Therefore, this strategy is inadequate.
The results for $S_2$, $S_3$ and $S_4$, instead, indicate good performance and are very similar to each other.
The low values for the average RMSEs $E$ and relative errors $\Delta$ for all the considered regions of interest, together with their reasonably low standard deviations, indicate that the estimates are both accurate and precise. These values of $\Delta$ show that the estimates of $P_{rad}^{\,tot}$ and of $P_{rad}^{\,main}$ exhibit little bias, while the estimates of $P_{rad}^{\,core}$ and of $P_{rad}^{\,div}$ exhibit some positive bias that is, however, acceptably low. Moreover, the RMSEs are also low and similar for all the regions of interest: larger relative errors are to be expected for regions that radiate less, such as the core and divertor. The correlation coefficients $r$ are close to $1$ for all regions, confirming that the overall trend of the radiated powers is well captured.
The statistics for $n_\sigma$ show that $\widehat{P}_{rad}^{\,tomo}$ is typically within one or two standard deviations $\sigma$ from $\widehat{P}_{rad}\;$: meaningful credible intervals can be obtained as $\widehat{P}_{rad}\pm\sigma$ or $\widehat{P}_{rad}\pm2\sigma$. \emph{These results confirm that the technique described in Section \ref{method} provides accurate real-time estimates of the total, core, divertor and main chamber radiated power; the associated credible bounds enable real-time uncertainty quantification.}\\
The similar performance achieved by the channel selection strategies $S_2$, $S_3$ and $S_4$ show that the technique is robust to the failure of a few channels ($S_3$ vs $S_2$) and to the exclusion of a few potentially good channels ($S_4$ vs $S_3$). This demonstrates that tracking the health of the bolometer channels over time makes the technique sufficiently robust for routine deployment in tokamak operation. Indeed, as mentioned earlier, the unexpected and sudden failure of a large number of channels is rare. For $S_4$, in only $\approx 3\%$ of the discharges from the considered TCV campaign the technique would have depended on one or more ill-behaved channels. As the results for $S_3$ show that a few channel failures can typically be safely handled, the technique can be expected to work robustly in the vast majority of cases. Importantly, this achievement is not obtained for a single scenario but over a number of diverse and relevant magnetic configurations. In terms of computational time, real-time estimation is extremely cheap as it is based on a simple linear combination of bolometer measurements as by Eq\eqref{eq_coefficients}. Pre-computing the coefficients for all regions of interest only required $\approx10\,\mathrm{s}$ per FBT equilibrium on a workstation equipped with Intel Xeon E5-2680 CPUs, used for all the experiments described herein. This is amply compatible with inter-discharge deployment at TCV ($\approx10$ minutes), with further acceleration achievable with more modern processors.

\section{Summary, Conclusion and Next Steps}\label{conclusion}
In this work, we proposed a novel technique for real-time estimation of the power radiated from any user-defined plasma region of interest. The key observation enabling the development of this technique is that, for Bayesian approaches featuring a Gaussian posterior, the radiated power can be efficiently computed as a linear combination of tomographic measurements. We derived the expression of the coefficients of the linear combination and showed that the obtained estimate is mathematically equivalent to the traditional tomographic estimate. We discussed how to include magnetic information in the estimations by leveraging the output of plasma equilibrium solvers used during TCV discharge preparation. The technique was applied to a set of 50 TCV discharges, chosen as a representative set and featuring a range of magnetic configurations.
The study showed that the total, core, divertor and main chamber radiated power can be accurately estimated in real-time by the proposed technique.
The technique also provides Bayesian credible intervals for the estimated quantities, allowing real-time uncertainty quantification.
Finally, we demonstrated that monitoring the failure history of the bolometer channels enables robust routine deployment. Open access is provided to all the data and computational routines used in the context of this work.\\
The proposed technique is computationally cheap and easy to deploy in real-time, since it only employs simple sets of coefficients. In contrast to statistical approaches that learn a mapping from bolometry measurements to the quantity of interest, our technique does not require a training phase. Thus, it bypasses the delicate issue of identifying a suitable training set and results in greater flexibility and generality.
The coefficients are pre-computed based on the plasma shape programmed for a given discharge during its preparation. Therefore, by design, the technique intrinsically accounts for different magnetic configurations and plasma shapes. It can be used to simultaneously track the power radiated from multiple user-defined regions of interest. By suitably choosing the geometry matrix and boundary conditions, it can explicitly account for vessel walls and non-convex structures such as the baffles for divertor closure often installed in TCV \cite{reimerdes_initial_2021}. The coefficients can also be updated to account for the failure history of the detectors. Its flexibility and versatility, combined with its robustness, make the technique ideally suited for routine deployment in plasma operation.\\
In terms of future work, we plan to integrate the technique into the control system of TCV. Together with suitable actuators such as fueling and impurity puffing, this will enable feedback control of the radiated power.
In particular, we intend to demonstrate the possibility of monitoring and effectively controlling the total radiated power. We also plan to support divertor detachment control experiments by providing estimates of the divertor radiated power. We further envision applications to primary/secondary X-point control and MARFE tracking, which could rely on the definition of smaller, experiment-specific regions of interest.
Future studies could also investigate possible improvements of the technique. Real-time anomaly detection and replacement models could be used to further increase robustness against detector failures. Moreover, while improved plasma control is expected to bring the experimental LIUQE equilibria closer to the FBT ones, we could integrate into our technique the information provided by the real-time version of LIUQE (RT-LIUQE). This would improve the estimations of the power radiated from regions of interest such as the core or the divertor, in cases where disagreements between FBT and LIUQE can be large. This could be implemented by pre-computing the matrix $\mathbf{A}$ in Eq.\eqref{eq_coefficients}, then computing in real-time the coefficients $\boldsymbol{\beta}$ by summing only over the indices of $\mathbf{A}$ corresponding to pixels belonging to the RT-LIUQE-defined region of interest. Finally, comparative studies with other real-time capable techniques could also be conducted.

\section*{Acknowledgements}
This work was supported in part by the Swiss National Science Foundation. This work has been carried out within the framework of the EUROfusion Consortium, via the Euratom Research and Training Programme (Grant Agreement No 101052200 — EUROfusion) and funded by the Swiss State Secretariat for Education, Research and Innovation (SERI). Views and opinions expressed are however those of the author(s) only and do not necessarily reflect those of the European Union, the European Commission, or SERI. Neither the European Union nor the European Commission nor SERI can be held responsible for them.

\bibliographystyle{unsrt}
\typeout{}
\bibliography{tomo_fusion}

\newpage

\appendix
\section*{Appendix}

\renewcommand{\thesubsection}{\Alph{subsection}}

\subsection{Hyperparameter Tuning}
\label{appendix_hyperparam}
\begin{figure}[t]
\begin{subfigure}[b]{0.8\textwidth}
\!\!\includegraphics[]{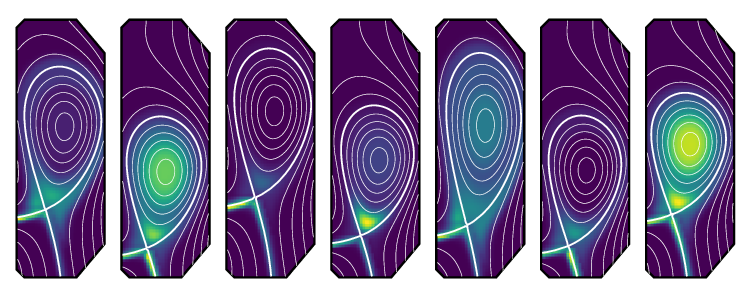}
\end{subfigure}
\hspace{1.3cm}
\begin{subfigure}[b]{0.0875\textwidth}
\centering
\raisebox{-4pt}{\includegraphics[width=\textwidth]{phantoms_cbar.eps}}
     \end{subfigure}
\captionsetup{justification=raggedright,singlelinecheck=true}
\caption{Synthetic dataset: example of model phantoms used for hyperparameter tuning.}
\label{fig_dataset}
\end{figure}

In this appendix, we discuss our hyperparameter tuning strategy, where by ``hyperparameters'' we collectively refer to the noise level $\eta$ in the likelihood and to the prior hyperparameters $\lambda, \alpha$. We show that, if the tomographic data $\mathbf{y}$ in Eq.\eqref{ip_data} are adequately normalized, a Gaussian posterior model akin to Eq.\eqref{eqs_posterior} with fixed values for $\eta, \lambda$ and $\alpha$ can be used to accurately estimate the radiated power.
We also show that enforcing a positivity constraint on the reconstructed emissivity is not necessary to obtain accurate results.
To do so, we carry out a study on a dataset $\mathcal{X}$ composed of $10^3$ physically realistic phantoms, to investigate the dependence of the tomographically estimated radiated power on these parameters.\\
\begin{figure}[t]
\begin{subfigure}[b]{0.85\textwidth}
\!\!\includegraphics[width=\textwidth]{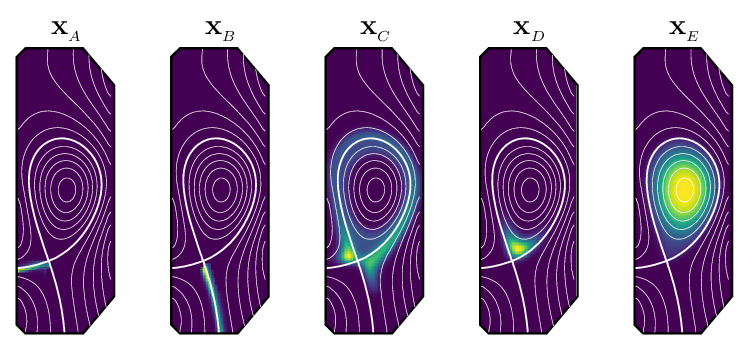}
\end{subfigure}
\hspace{0.6cm}
\begin{subfigure}[b]{0.0875\textwidth}
\centering
\raisebox{-4pt}{\includegraphics[width=\textwidth]{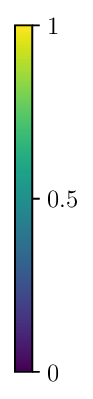}}
     \end{subfigure}
\captionsetup{justification=raggedright,singlelinecheck=true}
\caption{Phantom generation procedure: each phantom $\mathbf{x}$ is obtained as a linear combination of the five shown elements. Elements $\mathbf{x}_{_A}$, $\mathbf{x}_{_B}$, $\mathbf{x}_{_C}$ are obtained from a SOLPS phantom, by isolating its inner leg, outer leg and remaining radiation contributions, respectively; elements $\mathbf{x}_{_D}$ and $\mathbf{x}_{_E}$ are added to model X-point radiator and core radiation features, respectively.}
\label{fig_dataset_building_blocks}
\end{figure}
$\!\!\!$The synthetic emissivity profiles are generated by combining plasma boundary simulations with other physically realistic radiation features. The generation procedure is visually represented in Figure \ref{fig_dataset_building_blocks}.
We start from an emissivity profile generated by SOLPS \cite{wiesen_new_2015}, a leading plasma boundary simulation tool. In this profile, we isolate the radiation contributions from the inner divertor leg ($\mathbf{x}_{_A}$), outer divertor leg ($\mathbf{x}_{_B}$) and remaining plasma region ($\mathbf{x}_{_C}$). In addition, we create the radiation patterns $\mathbf{x}_{_D}$, modelling an X-point radiator feature, and $\mathbf{x}_{_E}$, modelling core radiation as a radially decaying profile whose peak coincides with the magnetic axis. The five radiation patterns $\mathbf{x}_{w}, \,w\!\in\!\{A,B,C,D,E\}$, are normalized to $1$ and defined on a uniform rectangular pixel grid of size $120\times41$. Composite phantoms $\mathbf{x}$ are then obtained by linear combination of the five described radiation patterns, as $\mathbf{x}=C\sum_{w} c_w \mathbf{x}_{w},\; \,w\!\in\!\{A,B,C,D,E\}$; both the linear combination coefficients $c_w$  and the scaling coefficient $C$ are randomly sampled from uniform distributions. With the adopted scalings, the total radiated power of the resulting phantoms satisfies $P_{rad}^{\,tot}\in[0\,\mathrm{MW},\;4\,\mathrm{MW}]$. To make the dataset further diverse, the magnetic equilibrium $\psi_\mathbf{x}$ associated to each phantom $\mathbf{x}$ is obtained by a randomized rescaling and shifting of the initial SOLPS equilibrium $\psi_{_{SOLPS}}$ in Figure \eqref{fig_dataset_building_blocks}. The radiation patterns $\mathbf{x}_{w}, \,w\!\in\!\{A,B,C,D,E\}$ are re-mapped to the new equilibrium $\psi_\mathbf{x}$ for phantom generation. In Figure \ref{fig_dataset}, we show some of the generated phantoms, normalized to $1$ for illustration purposes.\\
$\!\!\!$To study the sensitivity of the radiated power to the values of $\eta, \lambda$ and $\alpha$, we proceed as follows. For each phantom, we compute the associated tomographic data $\mathbf{y}$ by Eq.\eqref{ip_data}. In this equation, we model the acquisition by means of a volumetric geometry matrix $\mathbf{T}$, fully taking into account the three-dimensional geometry of the bolometer system. We always use this model throughout this work, both for tomographic reconstruction and coefficient computation. In addition, we model the noise $\boldsymbol{\eta}$ affecting the data to be signal-dependent, with $\eta_m^2=\eta_0^2+(0.05\,y_m)^2$, $m=1,\ldots,M$,  where $\eta_0$ represents background noise and $0.05\,y_m$ the signal-dependent component, corresponding to $5\%$ of the measured signal. The obtained data $\mathbf{y}$ are used as input for tomographic reconstruction.\\
To reconstruct a phantom $\mathbf{x}$ with tomographic data $\mathbf{y}$, we use a simplified model assuming signal-independent noise $\eta_m=\eta,$ $m=1,\ldots,M$. The mean and covariance matrix of the resulting Gaussian posterior are given by Eq.\eqref{eqs_posterior_flatnoise}, with $\psi$ the magnetic equilibrium associated to the phantom.
For reconstruction, we implement the normalization procedure discussed in Appendix \ref{normalization}. With this normalization, as we will now show, model hyperparameters such as $\eta, \lambda$ can be set to fixed values and treated as scale parameters.
Then, we compute the MAP $\mathbf{x}_{_{MAP}}$ from Eq.\eqref{MAP} for all combinations of $\eta, \lambda, \alpha$ with  $\eta\in\{10^{-2},2.5\cdot10^{-2},5\cdot10^{-2},7.5\cdot10^{-2},10^{-1}\}$, $\lambda\in\{10^{-3},10^{-2},10^{-1},10^{0},10^{1},10^{2},10^{3}\}$, $\alpha\in\{10^{-4},10^{-3},10^{-2},10^{-1},10^{0}\}$. We compute $\mathbf{x}_{_{MAP}}$ by means of the iterative proximal gradient descent (PGD) method described in \cite{hamm_tomography_2025}, setting to $0$ the value of all pixels lying outside the TCV vessel after each iteration. Instead, we do not enforce the positivity of the solution. From each $\mathbf{x}_{_{MAP}}$, we compute the tomographic estimate of the total and core radiated powers $\widehat{P}_{rad}^{\,tot,tomo}$ and $\widehat{P}_{rad}^{\,core,tomo}$ (with $P_{rad}^{\,core}$ defined as in Section \ref{results}), and the corresponding signed relative errors $\delta_{tot}=(\widehat{P}_{rad}^{\,tot,tomo}-P_{rad}^{\,tot})/P_{rad}^{\,tot}\,,\;$ $\delta_{core}=(\widehat{P}_{rad}^{\,core,tomo}-P_{rad}^{\,core})/P_{rad}^{\,core}$. Finally, we compute the mean ($\widetilde{\delta}$) and standard deviation ($s_\delta$) of these quantities over the set of $10^3$ phantoms $\mathcal{X}$.\\
\begin{figure}[t]
\begin{subfigure}[b]{0.45\textwidth}
\!\!\includegraphics[width=\textwidth]{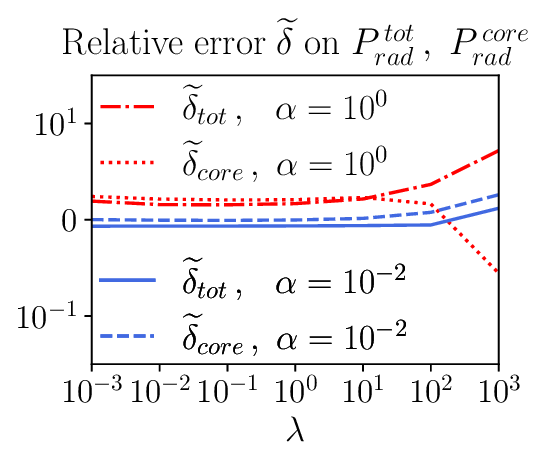}
\vspace{-0.3cm}
\caption{}
\label{figure_hyperparam_tuning}
\end{subfigure}
\hspace{-0.4cm}
\begin{subfigure}[b]{0.55\textwidth}
\renewcommand{\arraystretch}{2.15}
  \centering
\scalebox{0.98}{
  \begin{tabular}{|c|c|c|}
    \cline{1-3}
    $\alpha, \;\lambda$ & $\widetilde{\delta}_{tot}(\pm \, s_{\delta_{tot}})$ & $\widetilde{\delta}_{core}(\pm \,s_{\delta_{core}})$ \\ \hline
    $10^{-2},10^{1}$ & $-0.6\,(\pm2.8)\%$ & $0.1\,(\pm10.7)\%$ \\ \hline
    $10^0,10^{0}$ & $2.0\,(\pm3.4)\%$ & $2.2\,(\pm18.2)\%$ \\\hline
\end{tabular}
}
\vspace{0.75cm}
\caption{}
\label{table_hyperparam_tuning}
\end{subfigure}
\vspace{-0.3cm}
\captionsetup{justification=justified,singlelinecheck=true}
\caption{Hyperparameter tuning results over the set of model phantoms. Fig. \ref{figure_hyperparam_tuning}: average relative errors $\widetilde{\delta}_{tot},\,\widetilde{\delta}_{core}$ for the anisotropic model with $\alpha=10^{-2}$ and for the isotropic model with $\alpha=10^0$, as a function of the regularization parameter $\lambda$. Table \ref{table_hyperparam_tuning}: average value and standard deviation of the relative errors for the best performing anisotropic and isotropic models.}
\label{fig_reg_param_anis_param_tuning}
\end{figure}
$\!\!\!$By analyzing the results, we observe that the performance of the model is not highly sensitive to $\eta$ and $\lambda$, with excellent results achieved for a wide range of their values. Moreover, the results suggest that using anisotropic priors ($\alpha<1$) improves the accuracy of the estimates.
In Figure \ref{fig_reg_param_anis_param_tuning}, we compare the performance of the model corresponding to $(\eta,\alpha)=(2.5\cdot10^{-2}, 10^{-2})$, with that of the isotropic model $(\eta,\alpha)=(2.5\cdot10^{-2}, 10^{0})$, as a function of $\lambda$. From Figure \ref{figure_hyperparam_tuning}, we observe that the accuracy of both models stabilizes around a fixed value for small enough $\lambda$. Similar behavior is observed for the dependence on $\eta$ (not shown). By considering also the model performance in terms of the MSE $1/N\sum_{i=1}^N(\,(x_{_{MAP}})_i-x_i\,)^2$ (not shown), we select $\lambda=10^{1}$ and $\lambda=10^{0}$ for the anisotropic and isotropic models, respectively. Similar optimal values are obtained by using empirical Bayesian approaches such as those described in \cite{fernandez_vidal_maximum_2020}. Overall, these two are the best performing anisotropic and isotropic models. In Table \ref{table_hyperparam_tuning}, we report their performance for the estimation of $P_{rad}^{\,tot}$ and $P_{rad}^{\,core}$: the isotropic model exhibits a slight positive bias, while the anisotropic one has very little bias; moreover, the standard deviation of the isotropic model is significantly larger, especially in the case of $P_{rad}^{\,core}$. Thus, the results suggest that leveraging magnetic information makes the estimates more accurate and precise.\\
In conclusion, the hyperparameter study shows that $P_{rad}^{\,tot}$ and $P_{rad}^{\,core}$ can be reliably estimated by the discussed simplified model with $(\eta,\lambda,\alpha)=(2.5\cdot10^{-2}, 10^{1},10^{-2})$. The same applies to $P_{rad}^{\,div}$ and $P_{rad}^{\,main}$. The analysis confirms that, as remarked in \cite{hamm_tomography_2025, ingesson_l_c_comparison_1999, david_optimization_2021}, the power radiated from large regions of interest is much less sensitive to the hyperparameters than the single pixel values of the two-dimensional tomographic inversion. In Section \ref{results}, the selected model is used to compute the reference values $\widehat{P}_{rad}^{\,tomo}$. The real-time estimates $\widehat{P}_{rad}$ are obtained by a model with the same hyperparameters. For the computation of the coefficients $\boldsymbol{\beta}$, to approximate the behavior of the PGD algorithm used for $\widehat{P}_{rad}^{\,tomo}$, only the pixels lying within the TCV vessel are considered and no-flux boundary conditions are imposed in correspondence of the TCV wall.

\subsection{Normalization Procedure}
\label{normalization}
We describe here the normalization procedure followed in this work. First, we normalize $\mathbf{T}$ to 1, defining $\widetilde{\mathbf{T}}=\mathbf{T}/\max(\mathbf{T})$. Dividing also $\mathbf{y}$ by the same normalization factor we obtain $\widetilde{\mathbf{y}}=\mathbf{y}/\max(\mathbf{T})$. The inverse problem $\widetilde{\mathbf{T}}\mathbf{x}=\widetilde{\mathbf{y}}$ is equivalent to the original one $\mathbf{T}\mathbf{x}=\mathbf{y}$. However, this choice improves the numerical stability of the reconstruction algorithms, bringing all quantities closer to unity rather than working with very small values (for reference, the maximum value of the non-normalized $\mathbf{T}$ is around $10^{-11}$). Then, we further normalize $\widetilde{\mathbf{y}}$ to 1. Working with the input data $\overline{\mathbf{y}}=\widetilde{\mathbf{y}}/\max(\widetilde{\mathbf{y}})$ homogenizes the scale of the reconstruction problems, simplifying the handling of radiation profiles characterized by different radiation levels. The inverse problem to be solved now reads $\widetilde{\mathbf{T}}\overline{\mathbf{x}}=\overline{\mathbf{y}}$. The model hyperparameters are tuned based on this new problem. Operating with properly normalized versions of the geometry matrix and tomographic data simplifies the tuning and increases the robustness of the resulting reconstruction model. Multiplying the obtained solution $\overline{\mathbf{x}}$ by the normalizing factor $\widetilde{y}_{max}\coloneqq\max(\widetilde{\mathbf{y}})$ undoes the effects of the performed scaling, providing the solution $\mathbf{x}$ of the original problem.
Indeed, let $\overline{\mathbf{x}}\vert\overline{\mathbf{y}}\sim\mathcal{N}(\widetilde{\mathbf{A}}\overline{\mathbf{y}},\;\widetilde{\boldsymbol{\Sigma}}_{post})$ be the posterior associated to the normalized inverse problem $\widetilde{\mathbf{T}}\overline{\mathbf{x}}=\overline{\mathbf{y}}$. This coincides with the posterior discussed in Section \ref{tomography} (see Eq.\eqref{eqs_posterior}), with $\widetilde{\mathbf{A}}$ and $\widetilde{\boldsymbol{\Sigma}}_{post}$ obtained by replacing $\mathbf{T}$ with the normalized geometry matrix $\widetilde{\mathbf{T}}$ in the expressions of $\mathbf{A}, \boldsymbol{\Sigma}_{post}$.
Hence, since $\mathbf{x}=\widetilde{y}_{max}\,\overline{\mathbf{x}}$, it holds $\mathbf{x}\vert\overline{\mathbf{y}}\sim\widetilde{y}_{max}\,\mathcal{N}(\widetilde{\mathbf{A}}\overline{\mathbf{y}},\;\widetilde{\boldsymbol{\Sigma}}_{post})$, which implies $\mathbf{x}\vert\overline{\mathbf{y}}\sim\mathcal{N}\big(\widetilde{\mathbf{A}}(\widetilde{y}_{max}\,\overline{\mathbf{y}}),\;\widetilde{y}_{max}^{\,2}\,\widetilde{\boldsymbol{\Sigma}}_{post}\big)$ due to basic Gaussian distribution properties. Finally, as $\widetilde{\mathbf{y}}=\widetilde{y}_{max}\,\overline{\mathbf{y}}$, we have $\mathbf{x}\vert\widetilde{\mathbf{y}}\sim\mathcal{N}\big(\widetilde{\mathbf{A}}\widetilde{\mathbf{y}},\;\widetilde{y}_{max}^{\,2}\,\widetilde{\boldsymbol{\Sigma}}_{post}\big)$. The latter is precisely the model we are interested in, because of the already discussed equivalence of the problem $\widetilde{\mathbf{T}}\mathbf{x}=\widetilde{\mathbf{y}}$ to the original one.
Analogously, for the total radiated power expression derived in Section \ref{method}, if the normalized model yields $\overline{P}_{rad,\,post}^{\,tot}\sim \mathcal{N}( \mathbf{b}^T\,\overline{\boldsymbol{\mu}}_{post}, \,\mathbf{b}^T\,\widetilde{\mathbf{\Sigma}}_{post}\mathbf{b})$, the sought posterior distribution is $P_{rad,\,post}^{\,tot}\sim \mathcal{N}( \mathbf{b}^T\,\boldsymbol{\mu}_{post}, \,\widetilde{y}_{max}^{\,2}\,\mathbf{b}^T\,\widetilde{\mathbf{\Sigma}}_{post}\mathbf{b})$. Therefore, ultimately, all computations can be performed with the normalized model arising from $\widetilde{\mathbf{T}}\overline{\mathbf{x}}=\overline{\mathbf{y}}$; the obtained means and variances are then multiplied by a factor $\widetilde{y}_{max}$ and  $\widetilde{y}_{max}^{\,2}$, respectively.\\
For real-time estimation based on the linear combination in Eq.\eqref{eq_coefficients},
the sought $\widehat{P}_{rad}^{\,tot}$ is obtained directly from the data $\widetilde{\mathbf{y}}$, without normalizing them to $1$. Indeed, the coefficients $\boldsymbol{\beta}$ in Eq.\eqref{eq_beta} are not affected by the normalization of the data. Thus, the normalization can be neglected: $\widetilde{y}_{max}\sum_{j=1}^M\beta_jy_j/\widetilde{y}_{max}=\sum_{j=1}^M\beta_jy_j\,$. The normalization needs instead to be taken into account in the expression of the variance. If $\widetilde{\sigma}_{tot}^2=\mathbf{b}^T\,\widetilde{\boldsymbol{\Sigma}}_{post}\mathbf{b}$, the sought variance is given by $\sigma_{tot}^2=\widetilde{y}_{max}^{\,2}\,\mathbf{b}^T\,\widetilde{\boldsymbol{\Sigma}}_{post}\mathbf{b}$. Therefore, the predicted posterior variance depends on the maximum value of the $M$ tomographic data measured at a given time. The standard deviation $\sigma_{tot}$ can be used for real-time uncertainty quantification.
The same considerations apply to the estimation of the power radiated from other plasma regions (core, divertor, main chamber, $\ldots$).

\subsection{Plasma Discharge Numbers}
\label{shot_numbers}
In this appendix, we report the numbers of the 50 TCV discharges used to validate the technique in Section \ref{results}, subdivided according to their magnetic configuration. To avoid bias in the selection, the discharge numbers to be analyzed were selected purely based on their magnetic configuration, before carrying out the analysis presented in Section \ref{results}. They were selected in an attempt to create a diverse and representative dataset. For each magnetic configuration, we select plasma discharges based on different references, to increase variability in the plasma shape. Moreover, always with the goal of increasing the diversity of the chosen dataset, we include both L-mode and H-mode plasmas, discharges with and without ECRH heating, with and without gas injection. Below, the list of the used discharges.\\
Lower Single Null: 85184, 85185, 85269, 85270, 86554, 86572, 86605, 86735, 86850, 87081.\\
Upper Single Null: 85432, 85434, 86210, 86462, 86585, 86591, 86838, 86839, 87063, 87076.\\
Negative Triangularity: 84762, 84764, 85138, 85997, 86063, 86088, 86089, 86289, 86436, 86744.\\
X-point Target: 85199, 85350, 85353, 85357, 85815, 85816, 85818, 85819, 85820, 87080.\\
Long-Legged: 85166, 85174, 85192, 85194, 85439, 85487, 85489, 86015, 86117, 86985.

\end{document}